\documentclass[runningheads]{llncs}
\usepackage[T1]{fontenc}

\usepackage{graphicx}
\usepackage{booktabs}
\usepackage{amsmath}
\usepackage{amssymb}
\usepackage{xcolor}
\usepackage{CJK}
\usepackage{multirow}
\usepackage{makecell}
\usepackage{array}
\usepackage{makecell}
\usepackage{orcidlink}
\usepackage{marvosym}
\newcommand{\corr}{\textrm{\Letter}}

\begin{document}

\title{CUHK-EE Systems for the vTAD Challenge at NCMMSC 2025}

\author{Aemon Yat Fei Chiu\inst{1}\orcidlink{0009-0005-9672-8147} \and
Jingyu Li\inst{2}\orcidlink{0000-0001-5617-7014} \and
Yusheng Tian\inst{1}\orcidlink{0009-0008-6534-4943} \and
Guangyan Zhang\inst{2}\orcidlink{0000-0002-8640-8933} \and
Tan Lee\inst{1}\textsuperscript{(\corr)}\orcidlink{0000-0002-7089-3436}
}

\authorrunning{A. Y. F. Chiu et al.}

\institute{Department of Electronic Engineering, The Chinese University of Hong Kong, Hong Kong \\
\and
Independent Researcher \\
\email{\{aemon.yf.chiu,lijingyu0125,ystian0617,gyzhang\}@link.cuhk.edu.hk, tanlee@ee.cuhk.edu.hk}
}

\maketitle

\begin{abstract}

This paper presents the Voice Timbre Attribute Detection (vTAD) systems developed by the Digital Signal Processing \& Speech Technology Laboratory (DSP\&STL) of the Department of Electronic Engineering (EE) at The Chinese University of Hong Kong (CUHK) for the 20th National Conference on Human-Computer Speech Communication (NCMMSC 2025) vTAD Challenge. The proposed systems leverage WavLM-Large embeddings with attentive statistical pooling (ASTP) to extract robust speaker representations, followed by two variants of Diff-Net, i.e., Feed-Forward Neural Network (FFN) and Squeeze-and-Excitation-enhanced Residual FFN (SE-ResFFN), to compare timbre attribute intensities between utterance pairs. Experimental results demonstrate that the WavLM-Large+FFN system generalises better to unseen speakers, achieving 77.96\% accuracy and 21.79\% equal error rate (EER), while the WavLM-Large+SE-ResFFN model excels in the `Seen' setting with 94.42\% accuracy and 5.49\% EER. These findings highlight a trade-off between model complexity and generalisation, and underscore the importance of architectural choices in fine-grained speaker modelling. Our analysis also reveals the impact of speaker identity, annotation subjectivity, and data imbalance on system performance, pointing to future directions for improving robustness and fairness in timbre attribute detection.


\end{abstract}

\section{Introduction}

The Voice Timbre Attribute Detection (vTAD) Challenge \cite{ref_vtad2025,ref_vtad}, held in the 20th National Conference on Human-Computer Speech Communication (NCMMSC 2025), focuses on automatically determining relative intensity differences in voice timbre attributes between pairs of speech utterances, based on a set of pre-defined descriptor. Voice timbre is defined by a set of sensory descriptors in various modalities, including auditory (e.g., hoarse, rich), visual (e.g., bright, dark), tactile (e.g., soft, hard), and physical (e.g., magnetic, transparent) perceptions.

The design of CUHK-EE systems entering the vTAD Challenge leverages a large-scale speech self-supervised learning (SSL) representation model, namely the WavLM-Large \cite{ref_wavlm}, to extract robust voice features from speech signals. The voice features are refined using attentive statistical pooling (ASTP) \cite{ref_astp} before being passed into a comparison network, referred to as the Diff-Net. Two architectural variants of Diff-Net are investigated: standard Feed-Forward Neural Network (FFN) and Deep Residual FFN enhanced with Squeeze-and-Excitation blocks (SE-ResFFN) \cite{ref_rvector,ref_resnet,ref_se}.

The most critical component in our systems is the integration of WavLM-Large with layer-wise ASTP, which consistently provides strong and robust speaker embeddings in evaluation tracks. Although different downstream Diff-Net structures exhibit different performance in the two tracks, the upstream WavLM-Large with ASTP module remains unchanged and effective in all scenarios, suggesting that the WavLM-Large representations with layer-wise ASTP aggregation is effective for capturing and comparing voice timbre attributes.

\vspace{-3mm}

\begin{figure}
\centering
\includegraphics[width=0.9\textwidth]{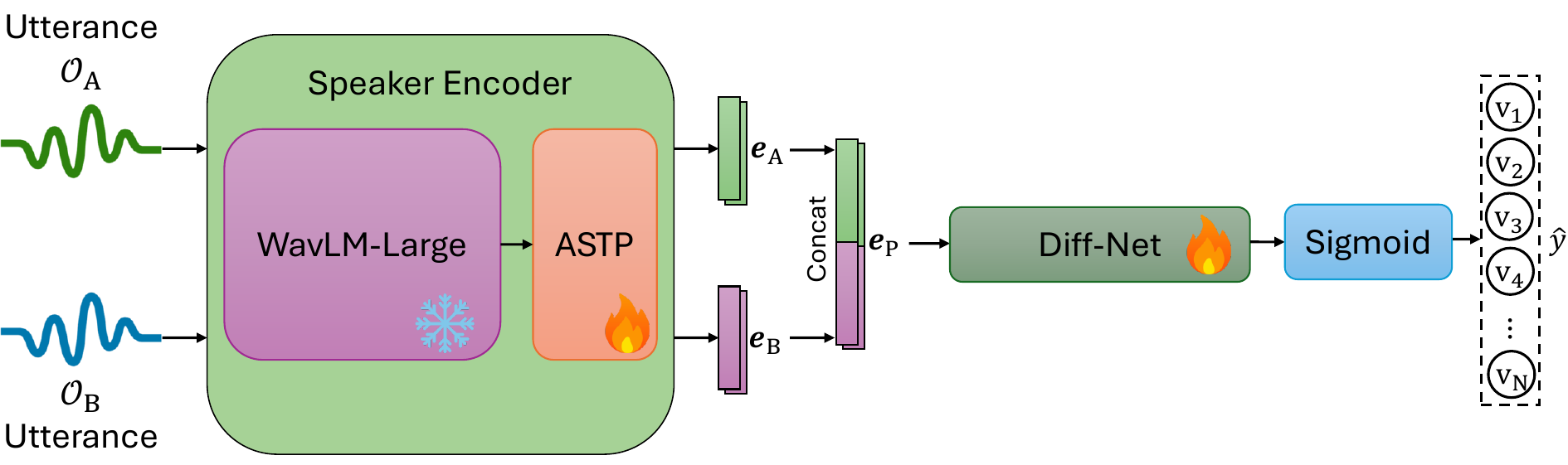}
\caption{The overall design concept of the systems.}
\label{overall}
\end{figure}

\vspace{-9mm}

\section{System Description}

\subsection{Audio Pre-Processing}

A silence removal procedure is applied to discard non-informative data from raw speech recordings. Silence is defined as regions where signal intensity is below 40 dB. The signal intensity is computed with a sliding window of 25 milliseconds (ms) long and 10 ms hop size. The process targets mainly the leading and trailing silence while preserving pauses in the middle of utterance, so that the natural rhythm and structure of speech are retained. To prevent excessive trimming, if the trimmed waveform is shorter than 100 ms, the silence removal step is bypassed to ensure the preservation of meaningful acoustic content.

\subsection{System Overview}

Figure \ref{overall} illustrates the overall architecture of the proposed system. It follows the baseline framework as described in \cite{ref_vtad2025,ref_vtad}, which consists of a speaker encoder and a comparison network.

Let the pair of input utterances be denoted as ${\mathcal O}_{\rm A}$ and ${\mathcal O}_{\rm B}$. The speaker encoder computes embeddings $\textbf{e}_{\rm A}$ and $\textbf{e}_{\rm B}$, which are concatenated to form a joint representation $\textbf{e}_{\rm P}$ to be passed to the Diff-Net module.

The Diff-Net module outputs a vector of dimension $N$. Each element of the output vector represents a specific timbre attribute. A sigmoid function is applied to the output of the Diff-Net to produce the prediction vector $\hat{\rm y}$. The $n$-th element of $\hat{\rm y}$ ($n = 1, 2, ..., 34$) gives the likelihood that ${\mathcal O}_{\rm B}$ has a stronger intensity than ${\mathcal O}_{\rm A}$ in the $n$-th timbre descriptor.

\vspace{-3mm}

\begin{figure}
\centering

\includegraphics[width=0.83\textwidth]{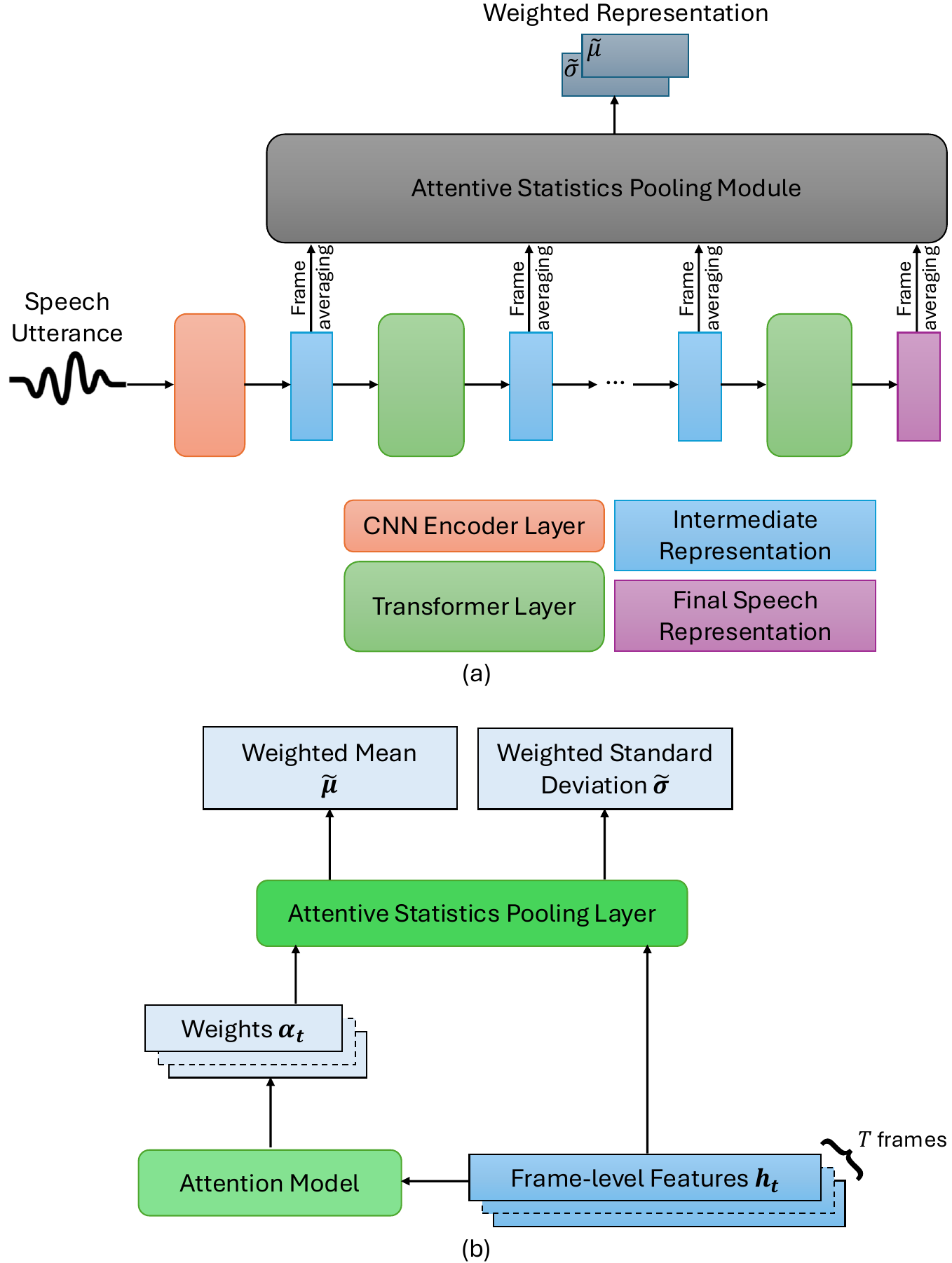}

\caption{(a) The WavLM-Large module with the adoption of ASTP for voice feature extraction. (b) The original ASTP machanism.}
\label{wavlm_astp}
\end{figure}

\vspace{-9mm}

\subsection{Speaker Encoder}

The WavLM-Large mode is leveraged as the speaker encoder for its superior ability to capture rich and hierarchical speech representations. WavLM was found to outperform traditional SV models and other SSL frameworks in generalisation and fine-grained timbre analysis \cite{ref_wavlm,ref_sfm}.

As shown in Figure \ref{wavlm_astp}, the speaker encoder integrates the WavLM-Large model \cite{ref_wavlm} with an ASTP module \cite{ref_astp}. The WavLM-Large model consists of a convolutional neural network (CNN) encoder layer followed by 24 stacked Transformer blocks. Each pre-processed utterance is fully fed into the model.

From the CNN encoder and each of the 24 Transformer blocks, 1024-dimensional frame-level intermediate representations can be obtained. Frame-level representations are averaged across frames. As a result, 25 layer-wise embeddings are obtained for each utterance.

To aggregate features from different layers, we employ ASTP instead of global pooling or selecting specific layers. The ASTP module dynamically weighs and summarises information from all layers, enabling the model to exploit the full depth of WavLM’s hierarchical structure for richer and more discriminative speaker embeddings.

The ASTP module was initially designed for frame-level aggregation in speaker verification. It is repurposed in our system to perform layer-wise feature aggregation. Specifically, eight attention heads are employed to compute weighted statistics (mean and standard deviation) across layers, yielding a 2048-dimension embedding for each utterance.

The embeddings from the two input utterances are concatenated to form a 4096-dimension feature vector, which serves as input to the Diff-Net. Two dropout layers with a dropout rate of 0.1 are incorporated in the ASTP module.

\vspace{-3mm}

\begin{figure}
\centering
\includegraphics[width=0.85\textwidth]{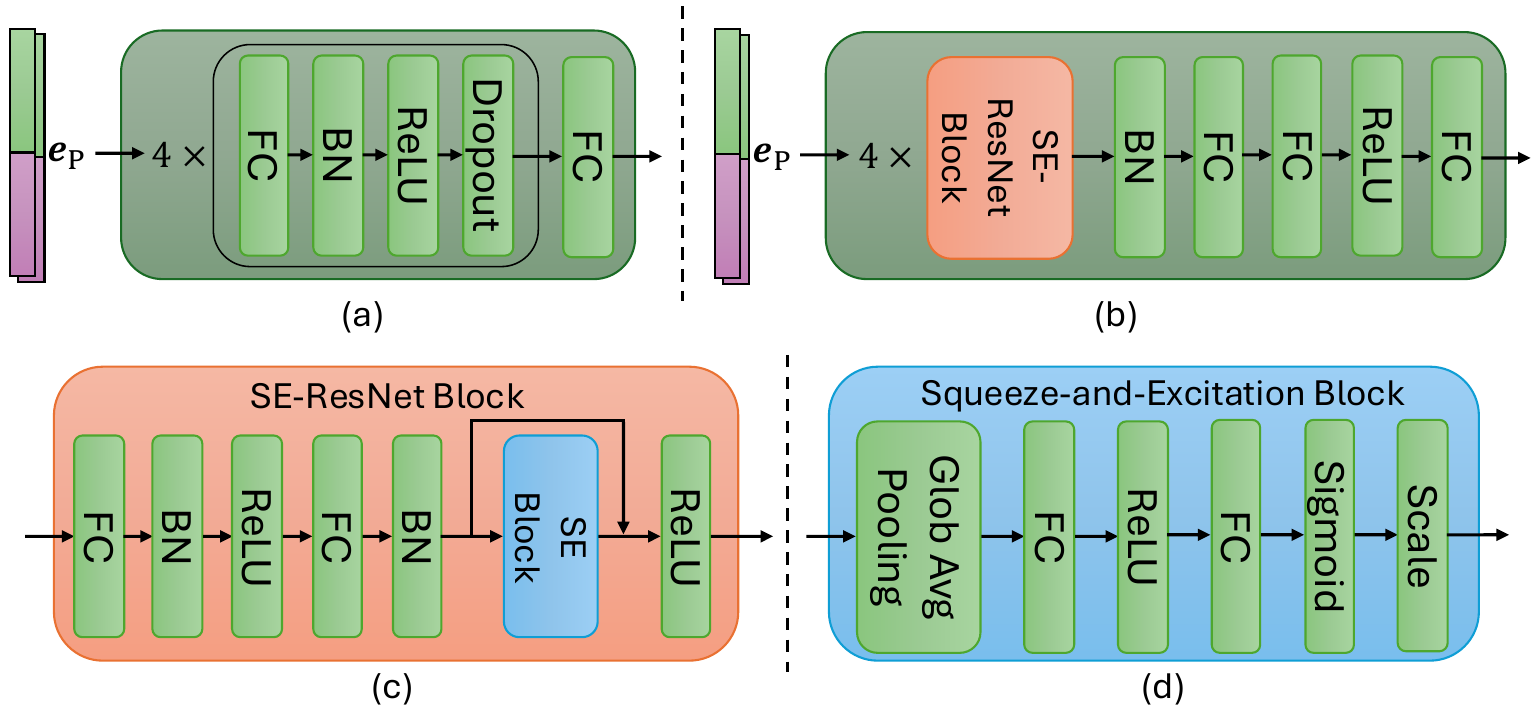}
\caption{(a) FFN-based Diff-Net. (b) SE-ResFFN-based Diff-Net. (c) An SE-ResNet block. (d) A squeeze-and-excitation block inside an SE-ResNet block.}
\label{diffnet-all}
\end{figure}

\vspace{-9mm}

\subsection{Diff-Net}

Figure \ref{diffnet-all} depicts the architectures of the two variants of Diff-Net investigated in our study.
The first variant, illustrated as in Figure \ref{diffnet-all} (a), is an enhanced version of the traditional FFN \cite{ref_vtad2025,ref_vtad}. It consists of four sequential FFN blocks with hidden dimensions of \textcolor{teal}{$[512,256,128,64]$}. Each block comprises a fully connected (FC) layer, followed by batch normalisation (BN), a ReLU activation function, and a dropout layer with a dropout rate of 0.3. After the final FFN block, an additional FC layer is applied to produce predictions for all 34 timbre attributes.

The second variant, SE-ResFFN, is inspired by the squeeze-and-excitation ResNet (SE-ResNet) architecture, which has shown strong performance in SV tasks \cite{ref_rvector,ref_se}. As illustrated in Figure \ref{diffnet-all} (b), (c), and (d), this model incorporates four SE-ResNet blocks with hidden dimensions of \textcolor{teal}{$[1024,1024,512,256]$}. Following these blocks, a BN layer is applied, and the resulting features are passed through two FC layers with hidden dimensions \textcolor{teal}{$[192,64]$}, interleaved with a ReLU activation. A final FC layer then outputs the predictions for the 34 timbre attributes.

Both architectures conclude with a sigmoid activation function, which transforms the raw outputs into probability scores representing the predicted likelihood that the second utterance exhibits a stronger intensity for each timbre attribute.

\section{Experimental Settings}

\subsection{Datasets}

The VCTK-RVA dataset \cite{ref_rva} is used for training the downstream Diff-Net modules (i.e., FFN and SE-ResFFN) and performance evaluation while the core component of the speaker encoder, the pre-trained WavLM-Large, is frozen in our training procedures. VCTK-RVA provides the annotation of 17 timbre descriptors for male and female speakers, resulting in a total of 34 distinct timbre attributes. The dataset was partitioned into a training set and two evaluation sets (`Seen' and `Unseen'), following the official challenge protocol. The training set includes 78 speakers (29 male and 49 female) forming 3,408 annotated speaker pairs and 136,320 utterance pairs. For evaluation, the `Seen' set consists of 76 speakers who also appear in the training set but with different utterances, comprising 235 speaker pairs and 94,000 utterance pairs, while the `Unseen' set contains 23 entirely new speakers forming 229 speaker pairs and 91,600 utterance pairs. The descriptors characterise different aspects of perceived voice quality, such as bright, thin, coarse, magnetic, shrill, and husky. The annotations are derived from human judgment, reflect subjective interpretations of vocal timbre. The subjectivity introduces variability in the data, which poses challenges for model consistency and generalisation to unseen speakers and utterances.

\subsection{Training Settings}

Model training was conducted over 10 epochs with a batch size of 16, using the AdamW optimiser with a learning rate of $1e-4$ and a weight decay of 0.01. A cosine annealing scheduler was employed to adjust the learning rate dynamically throughout training. The loss function used was binary cross-entropy with sample-wise reduction, and model validation was performed after each epoch. To ensure reproducibility, all experiments were initialised with a fixed random seed (42).

The proposed systems were evaluated under two distinct scenarios: `Unseen' and `Seen'. In the `Unseen' scenario, the speakers in the test set do not appear in the training set, thereby assessing the model’s generalisation capability. In the `Seen' scenario, the same set of speakers is used for both training and evaluation, but with non-overlapping utterances. Furthermore, for each target speaker, pairings with other individuals are uniquely assigned to either the training or evaluation set to prevent overlap.

Performance is measured using two standard metrics: accuracy (Acc) and equal error rate (EER). Accuracy reflects the proportion of correct predictions, while EER indicates the point at which false acceptance and false rejection rates are equal. Higher accuracy and lower EER values correspond to better system performance.

\vspace{-3mm}

\begin{table}[ht]
\centering
\caption{The results on the `Unseen' and `Seen' test sets.}
\renewcommand{\arraystretch}{2.2}
\setlength{\tabcolsep}{10pt}
\scalebox{0.875}{
\begin{tabular}{|>{\centering\arraybackslash}m{1.5cm}|>{\centering\arraybackslash}m{3.5cm}|>{\centering\arraybackslash}m{1.8cm}|>{\centering\arraybackslash}m{1.8cm}|}
\hline
\textbf{Test Set} & \textbf{Model} & \textbf{Acc (\%)} & \textbf{EER (\%)} \\
\hline
\multirow{2}{*}{Unseen} 
& \makecell[c]{WavLM-Large\\+\textbf{FFN}} & \textbf{77.96} & \textbf{21.79} \\
\cline{2-4}
& \makecell[c]{WavLM-Large\\+\textbf{SE-ResFFN}} & 74.90 & 25.17 \\
\hline
\multirow{2}{*}{Seen} 
& \makecell[c]{WavLM-Large\\+\textbf{FFN}} & 90.77 & 9.08 \\
\cline{2-4}
& \makecell[c]{WavLM-Large\\+\textbf{SE-ResFFN}} & \textbf{94.42} & \textbf{5.49} \\
\hline
\end{tabular}
}
\label{results}
\end{table}

\vspace{-9mm}

\section{Results and Analysis}

Table~\ref{results} summarises the overall performance of our systems under two evaluation scenarios: `Seen' and `Unseen'. As expected, both models perform substantially better on the `Seen' test set, where speakers are present in the training data, than on the `Unseen' set, which evaluates generalisation to novel speakers. For instance, the WavLM-Large+SE-ResFFN system achieves 94.42\% accuracy and 5.49\% EER on `Seen', but drops to 74.90\% accuracy and 25.17\% EER on `Unseen'. This gap highlights the difficulty of disentangling timbre attributes from speaker identity, suggesting potential over-fitting to speaker-specific patterns.

Interestingly, the WavLM-Large+FFN model outperforms WavLM-Large+SE-ResFFN in the `Unseen' setting, while WavLM-Large+SE-ResFFN excels in the `Seen' setting. This contrast may stem from the architectural complexity and inductive biases of the models. The SE-ResFFN-based Diff-Net, with its deeper layers and squeeze-and-excitation mechanisms, is better equipped to capture fine-grained speaker-specific patterns, which benefits performance when the test speakers are seen during training. However, this same specialisation may hinder generalisation to novel speakers, as the model may over-fit to specific traits rather than learning robust, speaker-invariant representations. In contrast, the simpler FFN-based Diff-Net architecture may generalise better due to its lower model complexity and reduced reliance on dynamic feature weighting, which may help it capture more speaker-invariant patterns and thus perform more robustly in the `Unseen' scenario. This observation suggests a trade-off between model expressiveness and generalisation, and highlights the importance of tailoring architectural choices to the target deployment context.

To further probe model robustness, we conducted small-scale experiments using speaker-disjoint training splits and observed considerable variation in results depending on the speaker composition. This, along with the inherent subjectivity in manual timbre labelling, suggests that model performance is sensitive to both speaker identity and annotation consistency. Additionally, the dataset \cite{ref_rva} exhibits imbalance in descriptor frequency and gender representation--some attributes are under-represented, and female speakers dominate the data, yet male speakers often yield better results. These factors collectively highlight the need for more balanced data and refined annotation practices to improve generalisation and fairness. Detailed results for the `Unseen' and `Seen' test sets, broken down by gender and descriptor, are listed in Table~\ref{tab:combined_results}.

\vspace{-3mm}

\begin{table}[ht]
\centering
\caption{Evaluation results of our proposed systems on the `Unseen' and `Seen' test sets. The rows `Average' are obtained by averaging the results across all the descriptors for each metric.}
\scalebox{0.875}{
\begin{tabular}{|c|c|c|c|c|c|c|c|}
\hline
\multirow{2}{*}{\textbf{Test set}} & \multirow{2}{*}{\textbf{Model}} & \multicolumn{3}{c|}{\textbf{Male}} & \multicolumn{3}{c|}{\textbf{Female}} \\
\cline{3-8}
& & \textbf{Attribute} & \textbf{Acc (\%)} & \textbf{EER (\%)} & \textbf{Attribute} & \textbf{Acc (\%)} & \textbf{EER (\%)} \\
\hline
\multirow{12}{*}{Unseen} & \multirow{6}{*}{\shortstack{WavLM\\-Large\\+\textbf{FFN}}}
& Bright (\begin{CJK*}{UTF8}{bsmi}明亮\end{CJK*}) & 69.12 & 31.82 & Bright (\begin{CJK*}{UTF8}{bsmi}明亮\end{CJK*}) & 58.56 & 41.43 \\
& & Thin (\begin{CJK*}{UTF8}{bsmi}單薄\end{CJK*}) & 73.71 & 24.67 & Thin (\begin{CJK*}{UTF8}{bsmi}單薄\end{CJK*}) & 55.57 & 44.01 \\
& & Low (\begin{CJK*}{UTF8}{bsmi}低沉\end{CJK*}) & 83.35 & 15.97 & Low (\begin{CJK*}{UTF8}{bsmi}低沉\end{CJK*}) & 71.45 & 28.13 \\
& & Magnetic (\begin{CJK*}{UTF8}{bsmi}磁性\end{CJK*}) & 94.94 & 5.35 & Coarse (\begin{CJK*}{UTF8}{bsmi}粗\end{CJK*}) & 89.85 & 10.06 \\
& & Pure (\begin{CJK*}{UTF8}{bsmi}乾淨\end{CJK*}) & 88.79 & 10.61 & Slim (\begin{CJK*}{UTF8}{bsmi}細\end{CJK*}) & 88.79 & 5.81 \\
\cline{3-5} \cline{6-8}
& & Average & 81.98 & 17.69 & Average & 73.94 & 25.89 \\
\cline{2-8}
 & \multirow{6}{*}{\shortstack{WavLM\\-Large\\+\textbf{SE-ResFFN}}}
& Bright (\begin{CJK*}{UTF8}{bsmi}明亮\end{CJK*}) & 65.49 & 32.52 & Bright (\begin{CJK*}{UTF8}{bsmi}明亮\end{CJK*}) & 49.33 & 48.27 \\
& & Thin (\begin{CJK*}{UTF8}{bsmi}單薄\end{CJK*}) & 74.03 & 26.47 & Thin (\begin{CJK*}{UTF8}{bsmi}單薄\end{CJK*}) & 48.65 & 52.62 \\
& & Low (\begin{CJK*}{UTF8}{bsmi}低沉\end{CJK*}) & 90.67 & 10.05 & Low (\begin{CJK*}{UTF8}{bsmi}低沉\end{CJK*}) & 70.02 & 31.71 \\
& & Magnetic (\begin{CJK*}{UTF8}{bsmi}磁性\end{CJK*}) & 80.31 & 19.80 & Coarse (\begin{CJK*}{UTF8}{bsmi}粗\end{CJK*}) & 88.65 & 12.24 \\
& & Pure (\begin{CJK*}{UTF8}{bsmi}乾淨\end{CJK*}) & 88.38 & 11.72 & Slim (\begin{CJK*}{UTF8}{bsmi}細\end{CJK*}) & 93.40 & 6.33 \\
\cline{3-5} \cline{6-8}
& & Average & 79.78 & 20.11 & Average & 70.01 & 30.23 \\
\hline
\multirow{12}{*}{Seen} & \multirow{6}{*}{\shortstack{WavLM\\-Large\\+\textbf{FFN}}}
& Bright (\begin{CJK*}{UTF8}{bsmi}明亮\end{CJK*}) & 93.38 & 7.03 & Bright (\begin{CJK*}{UTF8}{bsmi}明亮\end{CJK*}) & 85.06 & 14.97 \\
& & Thin (\begin{CJK*}{UTF8}{bsmi}單薄\end{CJK*}) & 92.85 & 6.53 & Thin (\begin{CJK*}{UTF8}{bsmi}單薄\end{CJK*}) & 90.14 & 9.34 \\
& & Low (\begin{CJK*}{UTF8}{bsmi}低沉\end{CJK*}) & 96.05 & 3.37 & Low (\begin{CJK*}{UTF8}{bsmi}低沉\end{CJK*}) & 93.26 & 6.63 \\
& & Magnetic (\begin{CJK*}{UTF8}{bsmi}磁性\end{CJK*}) & 95.25 & 4.80 & Coarse (\begin{CJK*}{UTF8}{bsmi}粗\end{CJK*}) & 86.06 & 13.80 \\
& & Pure (\begin{CJK*}{UTF8}{bsmi}乾淨\end{CJK*}) & 79.40 & 20.33 & Slim (\begin{CJK*}{UTF8}{bsmi}細\end{CJK*}) & 96.25 & 3.95 \\
\cline{3-5} \cline{6-8}
& & Average & 91.39 & 8.41 & Average & 90.15 & 9.74 \\
\cline{2-8}
& \multirow{6}{*}{\shortstack{WavLM\\-Large\\+\textbf{SE-ResFFN}}}
& Bright (\begin{CJK*}{UTF8}{bsmi}明亮\end{CJK*}) & 96.66 & 2.83 & Bright (\begin{CJK*}{UTF8}{bsmi}明亮\end{CJK*}) & 85.61 & 12.96 \\
& & Thin (\begin{CJK*}{UTF8}{bsmi}單薄\end{CJK*}) & 97.42 & 3.00 & Thin (\begin{CJK*}{UTF8}{bsmi}單薄\end{CJK*}) & 89.21 & 10.79 \\
& & Low (\begin{CJK*}{UTF8}{bsmi}低沉\end{CJK*}) & 98.10 & 1.30 & Low (\begin{CJK*}{UTF8}{bsmi}低沉\end{CJK*}) & 97.49 & 3.32 \\
& & Magnetic (\begin{CJK*}{UTF8}{bsmi}磁性\end{CJK*}) & 99.62 & 0.40 & Coarse (\begin{CJK*}{UTF8}{bsmi}粗\end{CJK*}) & 90.60 & 9.66 \\
& & Pure (\begin{CJK*}{UTF8}{bsmi}乾淨\end{CJK*}) & 91.88 & 8.47 & Slim (\begin{CJK*}{UTF8}{bsmi}細\end{CJK*}) & 97.55 & 2.20 \\
\cline{3-5} \cline{6-8}
& & Average & 96.74 & 3.20 & Average & 92.09 & 7.78 \\
\hline
\end{tabular}
}
\label{tab:combined_results}
\end{table}

\vspace{-9mm}

\section{Conclusion}
This paper presents the CUHK-EE systems submitted to the vTAD Challenge at NCMMSC 2025, designed to detect perceptual differences in voice timbre attributes through pairwise utterance comparison. Our approach integrates WavLM-Large embeddings with attentive statistical pooling for robust speaker representation, followed by two variants of Diff-Net, i.e., FFN and SE-ResFFN, for attribute intensity comparison. The systems achieved second place in the `Unseen' track and fourth place in the `Seen' track, demonstrating competitive performance across both generalisation and speaker-specific scenarios.

The results highlight the rich representational capacity of speech SSL models like WavLM, especially when paired with carefully designed downstream architectures. Notably, the FFN model showed stronger generalisation to novel speakers, while SE-ResFFN excelled in capturing fine-grained patterns among known speakers, suggesting a trade-off between model complexity and robustness. Our analysis also underscores the challenges posed by speaker variability, annotation subjectivity, and data imbalance, pointing to key areas for future improvement. These findings pave the way for further research in fine-grained speaker modelling and voice attribute disentanglement using self-supervised speech representations.

\end{document}